# Relativistic Coulomb problem for *Z* larger than 137

A. D. Alhaidari

*Saudi Center for Theoretical Physics, Jeddah, Saudi Arabia*
*Department of Physics, King Fahd University of Petroleum & Minerals, Dhahran 31261, Saudi Arabia*

We propose a relativistic one-parameter Hermitian theory for the Coulomb problem with an electric charge greater than 137. In the non-relativistic limit, the theory becomes identical to the Schrödinger-Coulomb problem for all *Z*. Moreover, it agrees with the Dirac-Coulomb problem to order $(\alpha Z)^2$, where $\alpha$ is the fine structure constant. The vacuum in the theory is stable and does not suffer from the "charged vacuum" problem for all *Z*. Moreover, transition between positive and negative energy states could be eliminated. The relativistic bound states energy spectrum and corresponding spinor wavefunctions are obtained.

**1. Introduction**

The problem of strong electric field in quantum electrodynamics (QED) has been the focus of renewed research interest for a long time. For a review see, for example, [1,2] and references therein. One of the most important physical effects in a strong time-dependent electric field is the dynamical electron-positron pair production from vacuum. On the other hand, for sufficiently strong static electric potential electron-positron pairs could, in principle, be created spontaneously. However, the process of static pair creation, which is predicted by QED [3], has yet to be confirmed unequivocally by experiment [4].

The Dirac equation gives a good description of the relativistic electron under the influence of various kinds of potential couplings. In the Dirac-Coulomb problem, the ground state energy of the electron in a hydrogen-like atom decreases as the point nuclear charge $-Ze$ increases [1,5]. The Sommerfeld's fine-structure energy spectrum formula indicates that the ground state of the electron becomes zero for $Z\alpha = 1$, where $\alpha = e^2/4\pi\epsilon_0 \approx 1/137$ is the fine structure constant [5]. On the other hand, for $Z > \alpha^{-1} \approx 137$ this energy becomes a complex number. That is, the Dirac Hamiltonian operator becomes non-Hermitian. Self-adjoint extension of the Hamiltonian is frequently achieved by taking into account the finite size of the nucleus [1,2,5]. For example, replacing the point nucleus by a uniform charged sphere of total charge $-Ze$ and finite radius of the order of few Fermis. As a result, the ground state energy regains reality for all *Z*, but decreases below zero as *Z* increases until it reaches $-mc^2$ at the critical charge $Z_{cr}$ [1,2]. Increasing Z further forces the ground state to dive into the negative energy (lower) continuum and changes its character from a bound state to a resonant state, called a supercritical resonance [1,6]. An initially vacant supercritical state decays into an electron-positron pair; a free positron and a bound electron. Thus, the vacuum, which was perturbed by the supercritical Coulomb potential, becomes charged. However, if the original bound state that became supercritical was fully occupied before diving, then no pairs are produced. Nonetheless, the charge of the electron gets embedded into the charge density generated by the vacuum polarization charge distribution. The vacuum will thus be carrying a net charge equal to the total charge of the electrons in supercritical resonant states. All remedies to the strong coupling scenario invoke concepts and employ tools outside the framework of one-particle relativistic quantum mechanics where the Dirac-Coulomb problem is originally formulated. Of course, it is well established that QED in lowest order results in the Dirac-Coulomb theory. However, being a perturbative quantum field theory with $\alpha Z$ as the perturbation parameter, QED might not properly handle the strong coupling region where $\alpha Z > 1$ and where the Dirac-Coulomb Hamiltonian becomes non-Hermitian.

In this work, we propose an alternative description of the relativistic electron in a strong static field generated by a point charge while avoiding the problem of a "charged vacuum". It is formulated within the theory of one-particle relativistic quantum mechanics and as such gives no direct implication on QED. It could, at best, be representing the lowest order limit of an alternative quantum field theory of electro-dynamics at strong coupling (e.g., a non-perturbative version of QED). Specifically, we are proposing a one-parameter theory based on the Dirac equation with coupling to the vector Coulomb potential and a "pseudo Coulomb potential" (to be defined below). We require that this theory agrees with the original Dirac-Coulomb problem to order $(\alpha Z)^2$. Moreover, in the nonrelativistic limit it must reproduce the Schrödinger-Coulomb problem for all *Z*. In the following section, we define the problem and present our approach to the solution. This work is an extension to, and/or departure from earlier work on the subject wherein a scalar Coulomb coupling is introduced in addition to the vector Coulomb coupling [7]. We show that there is a physical difference between the "pseudo Coulomb potential" introduced in this work and the scalar Coulomb potential. We give the new relativistic energy spectrum and corresponding spinor wavefunctions.

## 2. Dirac-Coulomb problem for *Z* > 137

The solution of the Dirac-Coulomb problem is defined as the solution of the Dirac equation for an electron (mass *m* and electric charge *e*) in the field of a static point charge. That is, the vector potential in the Dirac equation has zero space component, while the time component is the attractive Coulomb potential $V(r) = -\alpha Z/r$. Due to spherical symmetry, the equation separates into radial and angular components. The solution of the angular component is standard and is independent of *Z* [5]. In the conventional relativistic units ($\hbar = c = 1$), the Sommerfeld's fine-structure formula for the relativistic energy spectrum reads as follows [5]

$$\varepsilon_n^\kappa = \pm m \left[ 1 + \left( \frac{\alpha Z}{n + \sqrt{\kappa^2 - \alpha^2 Z^2}} \right)^2 \right]^{-1/2} \quad ; \quad n = 0, 1, 2, ..., \tag{1}$$

where $\kappa$ is the spin-orbit quantum number defined as $\kappa = \pm\left(j + \tfrac{1}{2}\right) = \pm 1, \pm 2, ...$ for $\ell = j \pm \tfrac{1}{2}$. For real solutions, the Dirac-Coulomb Hamiltonian is Hermitian and the energy spectrum must be real. Therefore, it is mandatory that the physical parameters satisfy $\alpha Z \leq |\kappa|$. Since $|\kappa| = 1, 2, ...$, then we must choose $Z \leq \alpha^{-1} \approx 137$. To overcome this problem for larger Z, we propose an *equivalent* relativistic Coulomb theory that carries a Hermitian representation for the case $Z > 137$. "Equivalence" is defined here as a relativistic Dirac theory that agrees with the original Dirac-Coulomb problem (for $Z \leq 137$) up to order $(\alpha Z)^2$ and whose nonrelativistic limit is identical to the Schrödinger-Coulomb problem for all Z. To do that, we proceed as follows.

In the units $\hbar = c = 1$ and in the standard representation of the Dirac matrices, the two-component radial equation with coupling to spherically symmetric scalar and vector potentials, where the space component of the later vanishes, reads as follows [5,7,8]

$$\begin{pmatrix} m+(V+W)-\varepsilon & \frac{\kappa}{r}-\frac{d}{dr} \\ \frac{\kappa}{r}+\frac{d}{dr} & -m+(V-W)-\varepsilon \end{pmatrix} \begin{pmatrix} \chi^+(r) \\ \chi^-(r) \end{pmatrix} = 0, \quad (2)$$

where $W(r)$ and $V(r)$ are the scalar and vector potentials, respectively. Now, we take V and W to be Coulomb-like; $V(r) = -\alpha v/r$ and $W(r) = -\alpha \mu/r$, where $\mu$ and $v$ are real parameters. We write Eq. (2) as $(H-\varepsilon)|\chi\rangle = 0$, where H is the 2×2 Hamiltonian matrix. Then, we apply to it the unitary transformation $e^{\frac{i}{2}\theta\sigma_2}(H-\varepsilon)e^{-\frac{i}{2}\theta\sigma_2}|\phi\rangle = 0$, where $\theta$ is a real constant parameter, $\sigma_2 = \begin{pmatrix} 0 & -i \\ i & 0 \end{pmatrix}$, and $|\phi\rangle = e^{\frac{i}{2}\theta\sigma_2}|\chi\rangle$. The requirement that $\theta$ satisfy the constraint $\mu\cos\theta - \frac{\kappa}{\alpha}\sin\theta = \pm v$, where $-\frac{\pi}{2} \leq \theta \leq +\frac{\pi}{2}$, takes Eq. (2) into

$$\begin{pmatrix} mC_\pm - \varepsilon - (1\pm 1)\frac{\alpha v}{r} & -mS_\pm + \frac{\gamma}{r} - \frac{d}{dr} \\ -mS_\pm + \frac{\gamma}{r} + \frac{d}{dr} & -mC_\pm - \varepsilon - (1\mp 1)\frac{\alpha v}{r} \end{pmatrix} \begin{pmatrix} \phi^+(r) \\ \phi^-(r) \end{pmatrix} = 0, \quad (3)$$

where $C = \cos\theta$, $S = \sin\theta$, and

$$\gamma = \kappa C + \alpha\mu S = \kappa\sqrt{1 + \left(\frac{\alpha}{\kappa}\right)^2(\mu^2 - v^2)}, \quad (4)$$

$$\begin{pmatrix} \phi^+ \\ \phi^- \end{pmatrix} = e^{\frac{i}{2}\theta\sigma_2}\begin{pmatrix} \chi^+ \\ \chi^- \end{pmatrix} = \begin{pmatrix} \cos\frac{\theta}{2} & \sin\frac{\theta}{2} \\ -\sin\frac{\theta}{2} & \cos\frac{\theta}{2} \end{pmatrix}\begin{pmatrix} \chi^+ \\ \chi^- \end{pmatrix}. \quad (5)$$

In the Appendix, we show the calculation details. The parameters are chosen such that the Dirac Hamiltonian in (3) is Hermitian. Now, the same square root that appears in $\gamma$ is also present in the rest of the parameters, $C_\pm$ and $S_\pm$. Therefore, reality dictates that $v^2 - \mu^2 \leq \alpha^{-2}$. Unlike formula (1) above; for any value of the vector potential parameter $v$ we can always choose $\mu$ to make this square root real. Equation (3) gives one radial spinor component in terms of the other as follows

$$\phi^{\mp}(r) = \frac{1}{\varepsilon \pm mC_\pm}\left(-mS_\pm + \frac{\gamma}{r} \pm \frac{d}{dr}\right)\phi^\pm(r), \quad (6)$$

Using this back in Eq. (3) results is the second order radial differential equation

$$\left[-\frac{d^2}{dr^2} + \frac{\gamma(\gamma \pm 1)}{r^2} - 2\alpha\frac{\varepsilon v + m\mu}{r} - (\varepsilon^2 - m^2)\right]\phi^\pm(r) = 0, \quad (7)$$

which is Schrödinger-like. Equation (6) is referred to as the "kinetic balance relation", which is not valid for $\varepsilon = \mp mC_\pm$. Now, since $1 \geq C_\pm \geq 0$, then the energy value $\varepsilon = \mp mC_\pm$ belongs to the negative/positive energy spectrum, respectively. Therefore, the top/bottom signs in Eqs. (6) and (7) correspond to positive/negative energy solutions, respectively. Since the two solution spaces are completely disconnected, we have to choose one of the two signs of the energy spectrum and obtain the corresponding solution, but not both. In what follows, we choose the top signs and obtain the positive energy solutions. The negative energy solutions are then obtained from these simply by applying the following map:

$$\phi^{\pm} \to \phi^{\mp},\ \varepsilon \to -\varepsilon,\ \kappa \to -\kappa,\ \nu \to -\nu,\ \mu \to \mu. \tag{8}$$

Note also that under this map: $C_{\pm} \to C_{\mp}$ and $S_{\pm} \to -S_{\mp}$.

We must now show that the nonrelativistic limit gives the correct Schrödinger-Coulomb problem defined by the solution of the following second order differential equation (written in the same units, $\hbar = c = 1$)

$$\left[ -\frac{d^2}{dr^2} + \frac{\ell(\ell+1)}{r^2} - 2m\frac{\alpha Z}{r} - 2mE \right]\psi(r) = 0, \tag{9}$$

where $\ell$ is the orbital angular momentum quantum number. Now, the nonrelativistic limit of Eq. (7) for positive energy (the top sign) is obtained by taking $\varepsilon \cong m + E$, where $|E| \ll m$. This results in the same Schrödinger equation (9) with $\ell = \begin{cases} \gamma & ,\gamma>0 \\ -\gamma-1 & ,\gamma<0 \end{cases}$ but with an electric charge $Z = \nu + \mu$. Thus, the question now is as follows: For $Z > 137$, can we find a pair of real parameters $\nu$ and $\mu$ such that $\nu + \mu = Z$ and $\nu^2 - \mu^2 \leq \alpha^{-2}$? The answer is definitely YES. For example, simply by choosing $\mu = \nu = \frac{1}{2}Z$, then $Z$ could take any desired value! Now, comes the question of *uniqueness*: Is there a unique and/or natural way to select a fixed value for one of the two parameters $\mu$ or $\nu$? We leave the answer to this question for later and note that with these coupling parameters the Dirac equation (2) becomes

$$\begin{pmatrix} m - \alpha\frac{Z}{r} - \varepsilon & \frac{\kappa}{r} - \frac{d}{dr} \\ \frac{\kappa}{r} + \frac{d}{dr} & -m - \alpha\frac{Z}{r} + 2\alpha\frac{\mu}{r} - \varepsilon \end{pmatrix} \begin{pmatrix} \chi^+(r) \\ \chi^-(r) \end{pmatrix} = 0. \tag{10}$$

Therefore, the *physical* content of the potential coupling in the Dirac Hamiltonian is a combination of the potential $\begin{pmatrix} -\alpha Z/r & 0 \\ 0 & -\alpha Z/r \end{pmatrix}$ and the potential $\begin{pmatrix} 0 & 0 \\ 0 & 2\alpha\mu/r \end{pmatrix}$. The former is the usual vector Coulomb potential while the latter is neither a scalar nor a vector. We call it the "*pseudo Coulomb potential*". It could be written as $\mathcal{V} - \mathcal{S}$, where $\mathcal{V} = \begin{pmatrix} \alpha\mu/r & 0 \\ 0 & \alpha\mu/r \end{pmatrix}$ is a vector potential and $\mathcal{S} = \begin{pmatrix} \alpha\mu/r & 0 \\ 0 & -\alpha\mu/r \end{pmatrix}$ is a scalar potential; both having equal magnitudes, $\alpha\mu$. We should also note that this constitutes a departure from earlier work on this problem, where a pure scalar Coulomb potential is added to the vector Coulomb potential (see, for example [7], and references therein). Moreover, the contribution of a pure scalar potential survives the non-relativistic limit, whereas this pseudo potential does not [8].

## 3. Energy spectrum

To obtain the bound states energy spectrum for this *equivalent* relativistic Coulomb problem, we first establish the parameter map between the relativistic equation (7) and the nonrelativistic equation (9). This map reads as follows:

$$\varepsilon > 0:\quad \psi \to \phi^+,\ Z \to \tfrac{\varepsilon}{m}\nu + \mu,\ E \to \tfrac{1}{2m}(\varepsilon^2 - m^2),\ \text{and}\ \ell \to \begin{cases} \gamma & ,\gamma>0 \\ -\gamma-1 & ,\gamma<0 \end{cases}. \tag{11a}$$

$$\varepsilon < 0: \quad \psi \to \phi^-, \; Z \to \tfrac{\varepsilon}{m}\nu + \mu, \; E \to \tfrac{1}{2m}(\varepsilon^2 - m^2), \text{ and } \ell \to \begin{cases} \gamma-1 & ;\gamma>0 \\ -\gamma & ;\gamma<0 \end{cases}. \quad (11b)$$

Using this map in the well-known nonrelativistic bound states energy spectrum formula, $E_{n\ell} = -mZ^2\alpha^2/2(n+\ell+1)^2$ with $n = 0,1,2,...$ [9], gives the following one-parameter ±tive energy spectrum

$$\varepsilon_n(\xi) = \pm m \left[1 + \left(\tfrac{\alpha Z}{n+|\gamma|}\right)^2 (1-\xi)^2\right]^{-1} \left[\pm \left(\tfrac{\alpha Z}{n+|\gamma|}\right)^2 \xi(\xi-1) + \sqrt{1 + \left(\tfrac{\alpha Z}{n+|\gamma|}\right)^2 (1-2\xi)}\right], \quad (12)$$

where the parameter $\xi = \mu/Z$ such that $2\xi \geq 1 - (\alpha Z)^{-2}$ and $\gamma = \kappa\sqrt{1 + \left(\tfrac{\alpha Z}{\kappa}\right)^2 (2\xi-1)}$. As a consistency check, we see that taking $\xi = 0$ gives the energy spectrum of the original Dirac-Coulomb problem (1). If we expand the energy spectrum (12) as a power series in $(\alpha Z)^2$ for $\alpha Z \ll 1$ and keep only the first order term, we obtain

$$\varepsilon_n(\xi) \approx \pm m\left[1 - \tfrac{1}{2}\left(\tfrac{\alpha Z}{n+|\gamma|}\right)^2\right]. \quad (13)$$

This agrees with the expansion of the Sommerfeld formula (1) up to order $(\alpha Z)^2$. Consequently, we choose to keep the parameter arbitrary since we have established the "*equivalence*" criterion mentioned above independently of $\xi$. The new energy spectrum formula (12) shows that the lowest positive energy bound state corresponds to $\gamma < 0$ and $n = 0$, where it becomes

$$\varepsilon_0(\xi) = m\left[\xi^2 + \left(\tfrac{\kappa}{\alpha Z}\right)^2\right]^{-1} \left[\xi(\xi-1) + \left(\tfrac{\kappa}{\alpha Z}\right)^2 \sqrt{1 + \left(\tfrac{\alpha Z}{\kappa}\right)^2 (2\xi-1)}\right] = mC_-. \quad (14)$$

Taking the limit $\alpha Z \to \infty$ shows that this minimum "positive" energy can never be less than $-m$ for any value of the parameter $\xi$ in the allowed range $2\xi \geq 1 - (\alpha Z)^{-2}$. Therefore, this theory does not suffer from the "charged vacuum" problem since the positive energy electron state can never become embedded into the negative energy continuum ($\varepsilon < -m$). Hence, the vacuum in this theory is stable. Moreover, if we require that positive and negative energy subspaces be disconnected, then the parameter $\xi$ will be restricted even further as $\xi \geq 1 - (\alpha Z)^{-1}$; otherwise, transition between positive and negative energy states can occur for large enough $\alpha Z$. In fact, this stronger condition on $\xi$ guarantees that the geometric cosine function satisfy $1 \geq C_\pm \geq 0$. Now, since the lowest positive energy is $mC_-$, then the highest negative energy is $-mC_+$ and the energy gap between the positive and negative energy spectra is

$$\Delta\varepsilon = m(C_+ + C_-) = \frac{2m\gamma/\kappa}{1 + (\alpha\xi Z/\kappa)^2}. \quad (15)$$

Figure 1 shows the energy spectrum (12) as a function of Z with no positive/negative energy transition $(\xi \geq 1 - (\alpha Z)^{-1})$ and with transition $(1 - (\alpha Z)^{-1} \geq \xi \geq \tfrac{1}{2} - \tfrac{1}{2}(\alpha Z)^{-2})$. In Fig. 2, we plot few of the lowest energy in the spectrum for a given Z and for a range of values of the parameter $\xi$ with $\xi \geq 1 - (\alpha Z)^{-1}$.

## 4. Spinor wavefunction

One method to obtain the two components of the radial spinor wavefunction is to use the parameter map (11). Applying this map on the non-relativistic Coulomb wave function [9] transforms it into the sought after eigenfunctions. However, we exploit here an alternative approach by proposing the following ansatz for the upper radial spinor component of the *positive* energy solutions

$$\phi_n^+(r) = A_n^\gamma (\lambda r)^\eta e^{-\zeta r} L_n^\rho(\lambda r), \tag{16}$$

where $\lambda > 0$, $\zeta > 0$, $\eta > 0$, $\rho > -1$, $A_n^\gamma$ is the normalization constant, and $L_n^\rho(z)$ is the associated Laguerre polynomials of degree $n = 0, 1, 2, \ldots$ Substituting (16) into (7) and using the differential equation of the Laguerre polynomial [10], we obtain

$$\zeta = \lambda/2, \quad \eta = \begin{cases} \gamma+1 & , \gamma > 0 \\ -\gamma & , \gamma < 0 \end{cases}, \text{ and } \rho = \pm(2\gamma+1) \text{ for } \pm\gamma > 0. \tag{17}$$

Moreover, we get the same energy spectrum formula (12) and $\lambda = \begin{cases} \lambda_{n+1} & , \gamma > 0 \\ \lambda_n & , \gamma < 0 \end{cases}$, where

$$\lambda_n = \frac{2\alpha Z}{n+|\gamma|}\left[\varepsilon_n(1-\xi) + m\xi\right]. \tag{18}$$

The wavefunction (16) becomes

$$\phi_n^+(r) = \begin{cases} A_n^\gamma (\lambda_{n+1} r)^{\gamma+1} e^{-\lambda_{n+1} r/2} L_n^{2\gamma+1}(\lambda_{n+1} r) & , \gamma > 0 \\ A_n^{-\gamma-1} (\lambda_n r)^{-\gamma} e^{-\lambda_n r/2} L_n^{-2\gamma-1}(\lambda_n r) & , \gamma < 0 \end{cases} \tag{19}$$

The lowest energy state at $\varepsilon_0$ corresponds to $\gamma < 0$ and $n = 0$. It reads as follows

$$\phi_0^+(r) = A_0^{-\gamma-1} (\lambda_0 r)^{-\gamma} e^{-\lambda_0 r/2}. \tag{20}$$

Substituting the upper spinor component (19) into the kinetic balance relation (6) with the top sign and using the differential and recursion properties of the Laguerre polynomials [10], we obtain the lower component as follows

$$\phi_n^-(r) = \begin{cases} \frac{\lambda_{n+1} A_n^\gamma}{\varepsilon_{n+1}+mC_+}(\lambda_{n+1} r)^\gamma e^{-\lambda_{n+1} r/2}\left[(n+2\gamma+1)L_n^{2\gamma}(\lambda_{n+1} r) - \left(\frac{mS_+}{\lambda_{n+1}}+\frac{1}{2}\right)(\lambda_{n+1} r)L_n^{2\gamma+1}(\lambda_{n+1} r)\right] & , \gamma > 0 \\ -\frac{\lambda_n A_n^{-\gamma-1}}{\varepsilon_n+mC_+}(\lambda_n r)^{-\gamma} e^{-\lambda_n r/2}\left[L_n^{-2\gamma}(\lambda_n r) + \left(\frac{mS_+}{\lambda_n}-\frac{1}{2}\right)L_n^{-2\gamma-1}(\lambda_n r)\right] & , \gamma < 0 \end{cases} \tag{21}$$

Again, the lowest energy state corresponds to $\varepsilon_0$ when $\gamma < 0$ and $n = 0$. It has the following lower spinor component

$$\phi_0^-(r) = -\frac{mS_+ + \lambda_0/2}{\Delta\varepsilon} A_0^{-\gamma-1} (\lambda_0 r)^{-\gamma} e^{-\lambda_0 r/2}. \tag{22}$$

Normalization of $\psi_0 = \begin{pmatrix} \phi_0^+ \\ \phi_0^- \end{pmatrix}$ gives the following normalization constant

$$A_0^{-\gamma-1} = \sqrt{\lambda_0/\Gamma(-2\gamma+1)}\left[1 + \left(\frac{mS_+ + \lambda_0/2}{\Delta\varepsilon}\right)^2\right]^{-1/2}. \tag{23}$$

Equations (19) and (21) show that the spinor wavefunction $\psi_n(r)$, whose components are $\phi_n^\pm(r)$, is associated with the energy $\varepsilon_n(\xi)$ for $\gamma < 0$ and with the energy $\varepsilon_{n+1}(\xi)$ for $\gamma > 0$.

The two radial components of the spinor wavefunction obtained above in (19) and (21) are for positive energies. To obtain the corresponding negative energy solutions, we apply on them the map (8). Note that $\nu \to -\nu$ and $\mu \to \mu$ imply that $Z \to (2\xi-1)Z$ and $\xi \to \frac{\xi}{2\xi-1}$. In Fig. 3, we plot the radial spinor components $\phi_n^\pm(r)$ for some of the lowest positive energy states and for $Z = 200$.

## 5. Conclusion

We added to the Dirac-Coulomb Hamiltonian a one-parameter "pseudo Coulomb potential". The result is a relativistic model for the Coulomb problem that maintained reality of the Hamiltonian of Hydrogen-like atoms for all $Z$. The lowest positive energy state does not dive into the vacuum (negative energy continuum) avoiding the problem of a charged vacuum. In fact, imposing a certain constraint on the parameter of the theory prevents transition between positive and negative energy states. Therefore, the space of solutions consists of two disconnected energy subspaces. This work embodies a departure from earlier work on the subject wherein a pure scalar Coulomb potential is introduced not the pseudo Coulomb potential presented here. In addition to the relativistic energy spectrum, we also obtained the corresponding spinor wavefunction.

**Acknowledgements:** This work is sponsored by the Saudi Center for Theoretical Physics. Partial support by King Fahd University of Petroleum and Minerals is highly appreciated.

**Appendix: Transformation of the Dirac equation (2) into equation (3)**

The unitary transformation $U(\theta) = \exp(\frac{i}{2}\theta\sigma_2)$ could be written as the following 2×2 matrix

$$U(\theta) = \begin{pmatrix} \cos\frac{\theta}{2} & \sin\frac{\theta}{2} \\ -\sin\frac{\theta}{2} & \cos\frac{\theta}{2} \end{pmatrix}. \tag{A1}$$

Applying this transformation to Eq. (2), we obtain

$$U \begin{pmatrix} m - \alpha\frac{\nu+\mu}{r} - \varepsilon & \frac{\kappa}{r} - \frac{d}{dr} \\ \frac{\kappa}{r} + \frac{d}{dr} & -m - \alpha\frac{\nu-\mu}{r} - \varepsilon \end{pmatrix} U^{-1} \begin{pmatrix} \phi^+ \\ \phi^- \end{pmatrix} = 0, \tag{A2}$$

where $U^{-1} = \begin{pmatrix} \cos\frac{\theta}{2} & -\sin\frac{\theta}{2} \\ \sin\frac{\theta}{2} & \cos\frac{\theta}{2} \end{pmatrix}$ and $\begin{pmatrix} \phi^+ \\ \phi^- \end{pmatrix} = U \begin{pmatrix} \chi^+ \\ \chi^- \end{pmatrix}$. We can write Eq. (A2) as follows

$$\left[ -\left(\alpha\frac{\nu}{r} + \varepsilon\right) U \begin{pmatrix} 1 & 0 \\ 0 & 1 \end{pmatrix} U^{-1} + \left(m - \alpha\frac{\mu}{r}\right) U \begin{pmatrix} 1 & 0 \\ 0 & -1 \end{pmatrix} U^{-1} \right. $$
$$\left. + \frac{\kappa}{r} U \begin{pmatrix} 0 & 1 \\ 1 & 0 \end{pmatrix} U^{-1} + \frac{d}{dr} U \begin{pmatrix} 0 & -1 \\ 1 & 0 \end{pmatrix} U^{-1} \right] \begin{pmatrix} \phi^+ \\ \phi^- \end{pmatrix} = 0 \tag{A3}$$

Using the following relations

$$U \begin{pmatrix} 1 & 0 \\ 0 & -1 \end{pmatrix} U^{-1} = \begin{pmatrix} C & -S \\ -S & -C \end{pmatrix}, \; U \begin{pmatrix} 0 & 1 \\ 1 & 0 \end{pmatrix} U^{-1} = \begin{pmatrix} S & C \\ C & -S \end{pmatrix}, \; U \begin{pmatrix} 0 & -1 \\ 1 & 0 \end{pmatrix} U^{-1} = \begin{pmatrix} 0 & -1 \\ 1 & 0 \end{pmatrix}, \tag{A4}$$

where $C = \cos\theta$ and $S = \sin\theta$, we obtain

$$\begin{pmatrix} mC - \alpha\frac{\nu}{r} + \frac{1}{r}(\kappa S - \alpha\mu C) - \varepsilon & -mS + \frac{1}{r}(\kappa C + \alpha\mu S) - \frac{d}{dr} \\ -mS + \frac{1}{r}(\kappa C + \alpha\mu S) + \frac{d}{dr} & -mC - \alpha\frac{\nu}{r} - \frac{1}{r}(\kappa S - \alpha\mu C) - \varepsilon \end{pmatrix} \begin{pmatrix} \phi^+ \\ \phi^- \end{pmatrix} = 0 \tag{A5}$$

This gives coupled first order differential equation for the two spinor components, $\phi^\pm$. By eliminating one component in terms of the other, we obtain a second order differential equation. We require that this becomes a Schrödinger-like equation (i.e., it contains no first order derivatives), which dictates that at lease one of the diagonal elements in (A5) be constant (i.e., independent of *r*). This means that we should choose the angle $\theta$ such that $-\alpha\frac{\nu}{r} \pm \frac{1}{r}(\kappa S - \alpha\mu C) = $ constant. It should be obvious (e.g., by equating coefficients of *r* in this condition) that this constant must be zero. Therefore, the Schrödinger-like requirement gives

$$\mu \cos\theta - \frac{\kappa}{\alpha}\sin\theta = \pm\nu. \tag{A6}$$

Writing $S = \pm\sqrt{1-C^2}$ gives $sign\sqrt{1-C^2} = \frac{\alpha}{\kappa}(\mu C \mp \nu)$, where *sign* is either + or − independent of the ± sign in front of *v*. Squaring both sides of this equation and solving the resulting quadratic equation for *C*, we obtain $C_\pm$ and $S_\pm$. Substituting (A6) in (A5) gives Eq. (3) with $\gamma = \kappa C + \alpha\mu S$.

**Figure Captions**

**Fig. 1**: The positive energy spectrum obtained from Eq. (12) as a function of $Z$ with allowed positive/negative energy transitions (dashed curves) and with no transitions (solid curves)

**Fig. 2**: The lowest positive energy in the spectrum given by Eq. (12) for $Z = 200$ and for a range of values of the parameter $\xi$ with $\xi \geq 1 - (\alpha Z)^{-1}$ (no positive/negative energy transitions)

**Fig. 3**: The radial spinor components $\phi_n^+$ (solid curve) and $\phi_n^-$ (dashed curve) for some of the lowest positive energy states and for $Z = 200$ with $\kappa = -1$ (Fig. 3a) and $\kappa = +1$ (Fig. 3b)

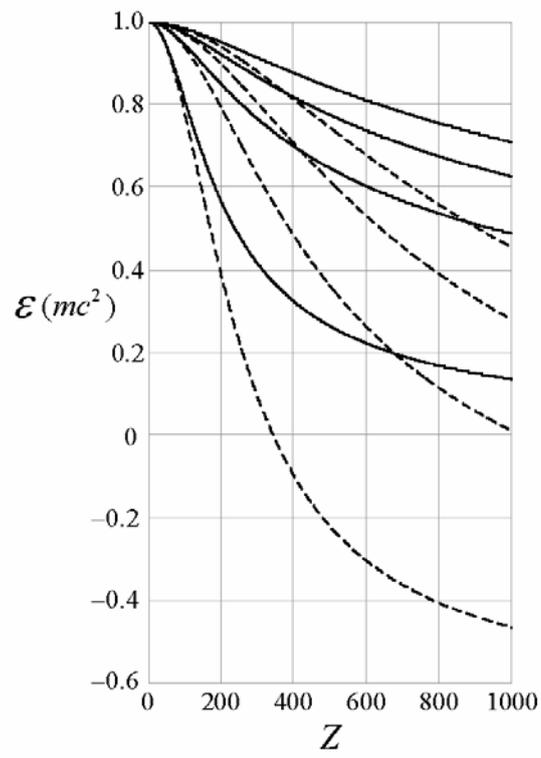

**Fig. 1**

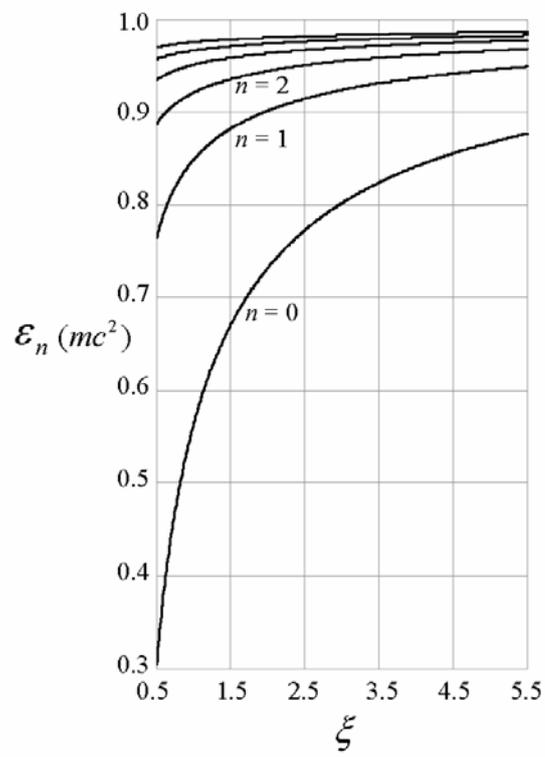

**Fig. 2**

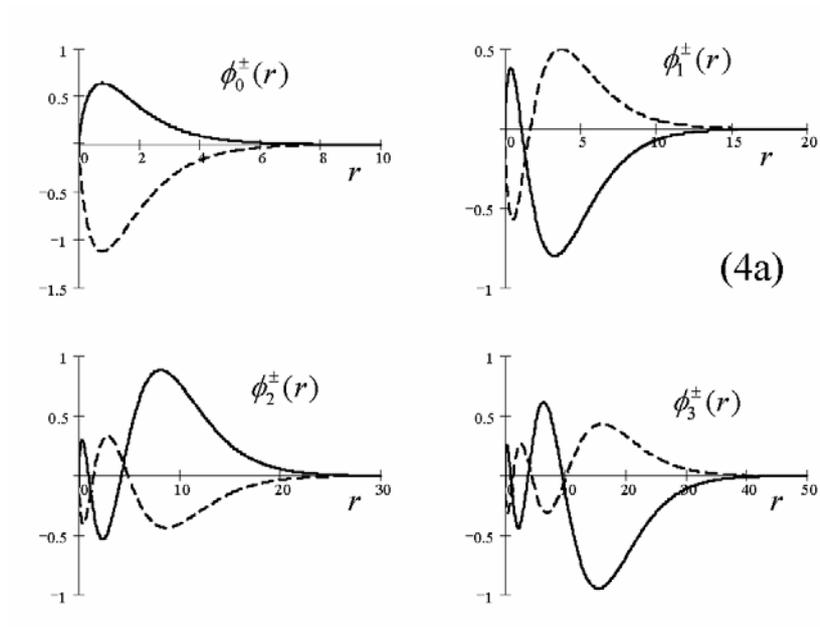

**Fig. 3a**

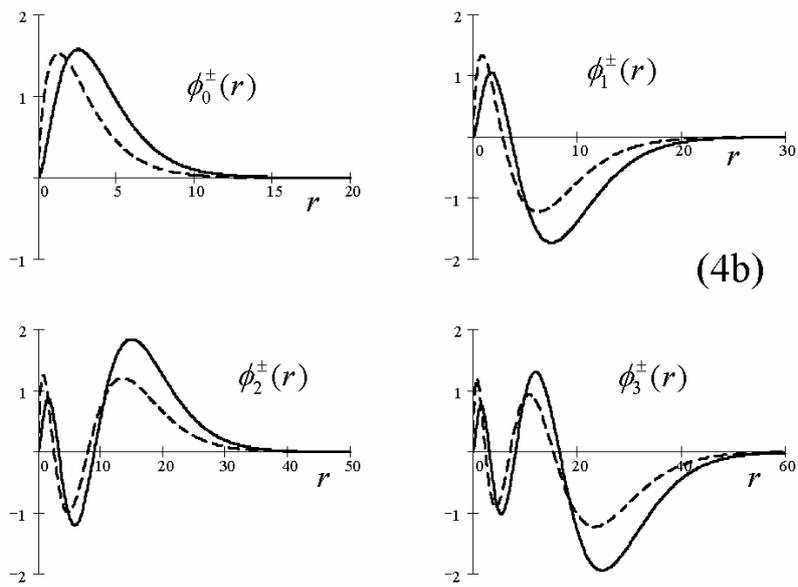

**Fig. 3b**